# Graphical Abstract

# Utilizing Sequential Information of General Lab-test Results and Diagnoses History for Differential Diagnosis of Dementia

Yizong Xing, Dhita Putri Pratama, Yuke Wang, Yufan Zhang, Brian E. Chapman

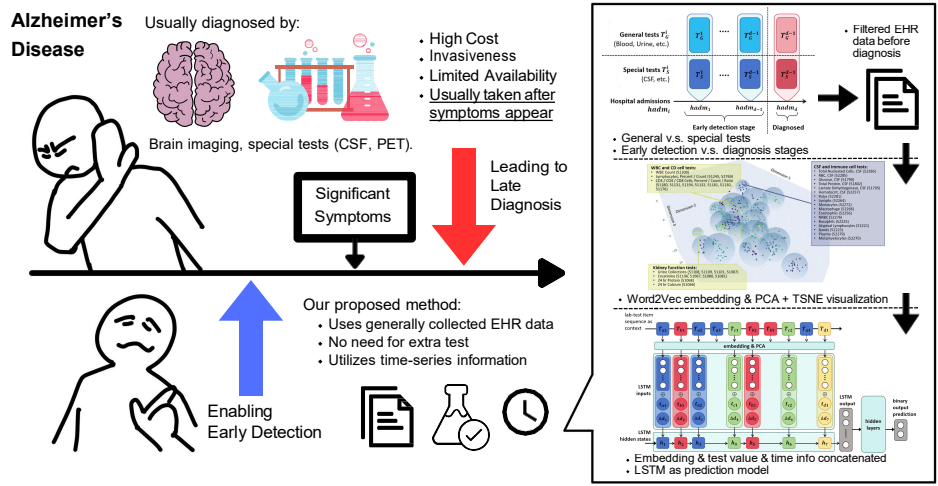

# Highlights

**Utilizing Sequential Information of General Lab-test Results and Diagnoses History for Differential Diagnosis of Dementia**

Yizong Xing, Dhita Putri Pratama, Yuke Wang, Yufan Zhang, Brian E. Chapman

- **Methodological Innovation**:
  - Treating lab-test sequences as "medical sentences", so as to apply Natural Language Processing techniques such as Word2Vec embeddings to capture representations and relationships of lab-tests, and innovatively integrated time-sequence neural networks for temporal modeling, enabling early AD detection using routine lab data alone.
  - Addressing data heterogeneity by proposing a unified framework to integrate highly sparse EHR data (eliminating reliance on AD-specific tests and traditional feature selection), validating the critical role of temporal information in AD prediction.

- **Key Findings**:
  - Clinical significance of temporal dynamics: Observed significant performance drops at 300 days pre-diagnosis, suggesting potential overlap between AD and other dementia subtypes during this phase.
  - Utility of general lab tests: Demonstrated that routine lab data (without specialized biomarkers) can enable cost-effective AD screening, promoting equitable healthcare access.

- **Technical Contributions**:
  - **Mixed vectorization strategy**: Developed a hybrid feature representation approach by combining Word2Vec embeddings for semantic encoding of lab tests, PCA and Kernel PCA for dimensionality reduction, and t-SNE for visualizing the underlying distribution of test patterns.

- **Robust benchmarking**: Performing systematic comparisons with logistic regression (LR), decision trees (DT) and random forests (RF), highlighting trade-offs between deep learning models and traditional methods in terms of precision, training efficiency, and interpretability.
- **Scalability and adaptability**: The model does not require disease-specific feature engineering and can be easily applied to other disease prediction tasks using structured EHR data.

- **Performance Strengths**:

  - **Competitive accuracy**: LSTM matched RF in most metrics while uncovering long-term temporal patterns inaccessible to tree-based models.
  - **Early detection validity**: Stable AUC (0.82–0.85) maintained within the 1-year prediagnosis window, confirming sensitivity to early pathological changes.

- **Implications and Challenges**

  - **Dataset limitations**: Exposed biases in public datasets (e.g., MIMIC-IV centric to the ICU, scarce prediagnosis samples) and advocated for longitudinal non-ICU AD cohorts.
  - **Model interpretability vs. complexity trade-off**: Although deep learning models (e.g., LSTM) offer strong performance, RF remain preferable in practical clinical scenarios due to their lower computational cost and better interpretability.
  - **Clinical translation**: Provides a methodological foundation for EHR-based AD screening tools, particularly in resource-limited settings.

# Utilizing Sequential Information of General Lab-test Results and Diagnoses History for Differential Diagnosis of Dementia

Yizong Xing[a,*], Dhita Putri Pratama[a,1], Yuke Wang[a,1], Yufan Zhang[a,1], Brian E. Chapman[a]

[a]*The University of Melbourne, 700 Swanston Street, Melbourne, 3053, VIC, Australia*

**Abstract**

**Objective:** Early diagnosis of Alzheimer's Disease (AD) is challenged by data variability, and reliance on specialized diagnostic tests with limited access. This study aims to explore whether routinely collected general laboratory tests and corresponding temporal information can be used to detect and differentiate early AD.

**Methods:** We propose a novel approach that treats sequential lab test records as structured time-series data. By applying word embedding techniques, we encode latent relationships among lab tests and utilize deep learning-based time-series models, including Long Short-Term Memory (LSTM) networks and Transformer architectures, to capture temporal patterns in patient records. The model is trained and evaluated on a public de-identified electronic health record (EHR) dataset.

**Results:** Our approach achieves competitive diagnostic performance and enhances early detection accuracy compared to traditional statistical models and baseline machine learning classifiers. The deep learning-based temporal modeling effectively distinguishes AD patients from non-AD controls. Performance metrics, including AUC, precision, and recall, indicate that general lab test histories contain predictive signals for AD diagnosis, even in the absence of specialized biomarkers.

**Conclusion:** The findings suggest that routinely collected laboratory tests can serve as a cost-effective and scalable alternative to early detec-tion of AD. This approach improves accessibility to early detection tools, particularly in resource-limited clinical settings, and lays the foundation for the integration of machine learning-based temporal models into real-world healthcare applications.

[*]Corresponding author: Yizong Xing (e-mail: yizongxing@unimelb.edu.au).
[1]Dhita Putri Pratama, Yuke Wang, and Yufan Zhang contributed equally to this work.

## 1. Introduction

Alzheimer's disease (AD), the leading cause of dementia, poses significant diagnostic challenges due to its prolonged progression and overlap of symptoms with other neurodegenerative disorders (e.g. Parkinson's disease, vascular dementia) and reversible conditions (e.g. vitamin deficiencies)[1]. These conditions share overlapping symptoms, making it difficult to distinguish AD from other dementia subtypes, particularly in the early stages when symptoms are mild and nonspecific. Although mild cognitive impairment (MCI) often precedes clinical dementia, current diagnostic frameworks struggle to differentiate between its underlying etiologies, making it difficult to initiate targeted interventions at an optimal stage[2, 3]. Inability to accurately classify early-stage dementia subtypes can lead to delays in treatment, suboptimal therapeutic strategies, and missed opportunities for clinical trials focusing on disease-modifying interventions[4]. Given the increasing prevalence of dementia worldwide, there is an urgent need for scalable, accessible, and cost-effective diagnostic approaches that can support early detection and differential diagnosis.

Existing methods are heavily based on specialized biomarkers such as cerebrospinal fluid (CSF) analysis, amyloid PET, and magnetic resonance imaging (MRI), which provide valuable information on disease pathology[5]. However, these tests are costly, invasive, and generally administered only after symptom escalation[6], when irreversible neurodegeneration has already occurred. Even models designed for early prediction frequently depend on such biomarkers[7], limiting their applicability in resource-constrained settings where these diagnostic tools are unavailable or infeasible for large-scale screening. Furthermore, variability in electronic health record (EHR) data— where some patients lack imaging scans, others have missing lab results, and ICD codes often suffer from inaccuracies—poses a major challenge to model generalizability[8]. The fragmented nature of the EHR data results in inconsistent availability of characteristics among patients, complicating the development of predictive models based on multimodal clinical inputs. Traditional imputation techniques can address small-scale missing values but struggle with widespread data sparsity, leading to suboptimal model performance and potential biases in patient classification[9].

To overcome these limitations, we propose a novel approach that utilizes longitudinal lab-test sequences, a ubiquitous but underutilized data source. Drawing inspiration from natural language processing (NLP), we first extract lab-test histories from EHR data, optimize the embedded features to reduce noise interference from ICD diagnoses, and improve the generalizability of the model using Principal Component Analysis (PCA[10]). Furthermore, we treat patient lab-test histories as text sequences and apply embedding techniques (e.g. Word2Vec[11, 12]) to capture similarities and differences in the data[13]. To address data heterogeneity and missing values, we incorporate the Long-Short-Term Memory (LSTM)[14] and Transformer[15] models, leveraging temporal information in laboratory test histories. These sequence models can extract diagnostic signals through dynamic associations, even in cases of sparse matrices or incomplete data, ensuring robust and accurate predictions. In general, the objective of our research is to:

- Capture the inter-feature sequential information of patients' biomarker lab-test histories using embedding technologies;

- Develop a classification model for AD utilizing the lab-tests' sequential information, which can work for both differential diagnosis (with special tests available) and early detection.

*Statement of Significance*

- **Problem:** Early detection of Alzheimer's disease (AD) is hindered by the limited availability of specialized diagnostic tests, high variability in patient data, and excessive reliance on single-type biomarkers. Existing diagnostic models often fail to take advantage of routinely collected general lab test histories for predictive insights.

- **What is Already Known:** Previous studies have explored machine learning models for AD diagnosis, relying primarily on neuroimaging, genetic biomarkers, or structured clinical assessments. However, the potential of general laboratory test sequences in early AD detection remains largely underexplored.

- **What this Paper Adds:** This study introduces a novel approach that models sequential lab test records as structured time-series data, applying word embedding techniques and deep learning-based temporal modeling (LSTM, Transformer) to extract predictive signals from routine lab tests. Our findings demonstrate that general lab test histories can improve diagnostic accuracy, enabling scalable and cost-effective AD screening.

- **Who Would Benefit from the New Knowledge in this Paper:** Clinicians, healthcare data scientists, and AI medical researchers seeking alternative, widely available, and cost-effective biomarkers for early detection and differential diagnosis of AD, particularly in resource-limited clinical settings.

## 2. Methodology

*2.1. Ethics Statement*

All data for this study was accessed from the MIMIC dataset[16] after completing appropriate ethics training and agreeing to the MIMIC Data Use Agreement[17]. For this study, we selected laboratory test histories relevant to Alzheimer's disease and related neurodegenerative conditions.

*2.2. Definitions and assumptions*

*2.2.1. "General" versus "special" lab-tests*

Since the objective of our work is to capture and utilize general lab-test information, a clear definition is essential to separate general tests ($T_G$) and AD-specific tests ($T_S$).

- Only laboratory items and fluids that are commonly tested in routine examinations of people are included (e.g. blood and urine tests) in $T_G$, excluding AD-specific tests ($T_S$) such as CSF and PET tests.

- Only lab-test biomarkers are included, excluding radiology reports obtained from brain imaging tests such as MRI.

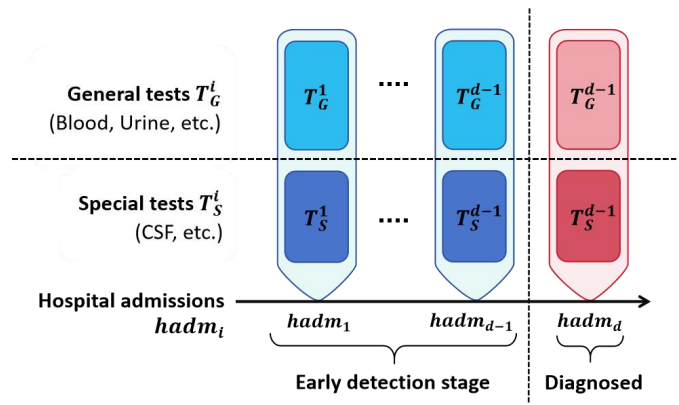

Figure 1: **Illustration of the definitions.** We would like to clarify the definitions and differences between 'general' and 'AD-specific' tests, as well as the 'early detection' and 'diagnosis' stages. Each hospital admission $hadm_i$ contains the general tests $T_G^i$ and the AD-specific tests $T_S^i$. **1.** $T_G$ include blood and urine tests, while $T_S$ include AD-or-dementia-specialized tests such as CSF tests. Since our model only deals with lab-tests, patient backgrounds (age, gender, education, economic status) and radiology results (e.g. brain MRI) are excluded in this study. **2.** $hadm_d$ represents the admission where the patient is diagnosed as AD or other dementia for the first time (corresponding ICD codes are found in the discharge table of this admission), while the early detection model considers only the early-stage ($S_E$) admissions before it. The diagnosis model considers the information from the complete stage $S_C$, which includes both $S_E$ and $hadm_d$.

*2.2.2. "Early detection" versus "diagnosis"*

As mentioned above, while there are specific tests widely used in the investigation of AD, they are generally acquired only after there is a significantly higher suspicion that the patient has AD. Thus, we believe that it is necessary to clearly distinguish the definitions between early detection and diagnosis tasks. We assume that the tests become "sufficient" only after an admission where the ICD code is given in the discharge information.

- **Early detection** occurs before sufficient lab-tests are available to confirm a diagnosis. We define admissions used for early detection as **early stage**, represented by $S_E$. We denote this task as $S_E \otimes T_G$, where $\otimes$ represents that $T_G$ within $S_E$ are being considered.

- **Diagnosis** occurs after a sufficient number of lab-tests have been conducted for diagnostic purposes. We define the admissions used for diagnosis as the complete stage, denoted $S_C$, including both $S_E$ and the admission where the ICD code for dementia is given. We denote this task by $S_C \otimes (T_G \cup T_S)$, where $\cup$ indicates the union of $T_G$ and $T_S$.

Table 1: ICD codes for dementia-related diseases. The table includes ICD-10 and ICD-9 codes for various forms of dementia, including Alzheimer's disease, vascular dementia, and others.

| Disease | ICD-10 Codes | ICD-9 Codes |
| --- | --- | --- |
| Alzheimer's Disease | G30.x | 331.0 |
| Vascular Dementia | F01.x | 290.4, 331.1 |
| Unspecified Dementia | F03.x | 290.0-290.3, 290.8-290.9 |
| Frontotemporal Dementia | F02.x | 331.1, 331.2 |
| Lewy Body Dementia | G31.8 | 331.82 |
| Senile Degeneration | R54 | 797 |
| Other Degenerative Diseases | G31.0, G31.1, G31.8, G31.9 | 331.0, 331.1, 331.2, 331.9 |

These definitions are suitable for our study, as we aim to develop a model based solely on general lab results, such as blood and urine tests. These types of test or investigations excluded here typically occur after a patient is suspected of potentially having AD, which is too late for early detection. In this research, the general tests $T_G$ are especially represented by blood-based tests since they are some of the most commonly taken tests and contain sufficient information to build the dataset. These definitions are further visualized and explained in Figure 1.

*2.3. Dementia cohort phenotyping*

In this research, we used MIMIC-III and MIMIC-IV datasets[18, 16], comprehensive collections of clinical, demographic and administrative hospital data for patients admitted to intensive care units (ICU). These data sets provide detailed electronic health records, including information on admissions, laboratory tests, medication administrations, and diagnostic codes, allowing robust analyses for critical care research.

The cohort was defined by the International Classification of Diseases (ICD) codes[19], a standardized representation of medical diagnoses and procedures. From a systematic review[20], we identified a set of 27 ICD-10 codes to define the cohort of dementia patients, as detailed in Table 1, and obtained corresponding ICD-9 codes from the MIMIC-III dataset. Patients with AD diagnosis were labeled True cases, while patients with other forms of dementia, as listed in Table 1, were labeled as False cases. We successfully collected 3947 patients from the MIMIC-IV dataset, including 1010 True cases (AD patients) and 2937 False cases (non-AD dementia patients). In addition, 1,168 AD patients were collected from the MIMIC-III dataset.

*2.4. Patient lab-test history extraction and embedding*

For each patient collected in the cohort, we extracted the complete set of all the laboratory tests (denoted as itemids) and result values, sorted by the discharge time of the diagnoses. Then, treating each patient's itemid sequence as the context and each itemid as a word, we embedded all the itemids into a 20-dimensional vector space using the *Word2Vec* algorithm[11, 12]. We then applied principle component analysis (PCA)[10] on the embedding dimensions to get better orthogonal representations. The new dimensions were sorted by their explained variance ratio, resulting in a revised 20-dimensional PC vector space. These new PC vectors for the itemids will later serve as inputs for the classification model. Additionally, we used the *t-SNE* algorithm[21] to reduce the dimension to 3 for better visualization.

*2.5. Classification models*

*2.5.1. Baseline models*

We used some classic binary classification models, including Logistic Regression (LR), Decision Tree (DT), and Random Forest (RF), as baselines. The input data to the baseline model was constructed by representing each patient as a vector of lab-test results. If a test was conducted multiple times within the same admission stage, the value used was the first recorded abnormal result (defined as values outside the reference ranges provided in MIMIC-IV). The vector dimension was determined by the total number of unique tests conducted across all patients in the cohort. Missing values were filled with the median of valid values from other patients. We further applied Lasso for feature selection, where the selected features rather than all the features would be used in the models.

*2.5.2. Our approach - LSTM-D*

As mentioned before, we employ an LSTM model[14] to process patients' laboratory test history data. The implementation is inspired by GRU-D[22], incorporating time information in the vectors to capture the temporal gaps between time steps.

In our approach, all tests from each admission are grouped together as one time step. For each time step:

1. A 20-dimensional PCA embedding vector ($T_{kj}$) encodes lab-tests by category ($k$) and their sequential order ($j$) within the category for the admission. In addition to evaluating PCA, we conducted experiments using 20 and 40 dimensional kernel PCA as an alternative dimensionality reduction technique to explore its potential in capturing non-linear structures in the data.

Table 2: Explanation of input vector dimensions and their represented symbols.

| Dim | Symb | Explanation |
| --- | --- | --- |
| 1-20 | $T_{kj}$ | PCA-processed embedding of lab-tests. Here, $k$ represents the test category (e.g., blood tests, imaging), and $j$ denotes the sequential order of the test within each category for the given admission. |
| 21 | $t_{kj}$ | Corresponding lab-test value. |
| 22 | $\Delta d_i$ | Time difference between the timestamp of this lab-test and the diagnosis time, calculated as $\Delta d_i = \log(\text{time(diag)} - \text{time}(T))$. This encodes the temporal relationship between tests and the diagnostic event. |

2. The resulting vector is then concatenated with the actual lab test value ($t_{kj}$) and the logarithmically scaled time difference ($\Delta d_i$), which represents the interval between the test's timestamp and the final diagnosis time. The log transform was applied to compress the wide dynamic range of raw time intervals, preventing large values from dominating gradient updates during model training.

The resulting 22-dimensional vectors are arranged chronologically based on the admission time to form the input matrix, as shown in Table 2. This chronological arrangement ensures the temporal sequence of tests is preserved, allowing the model to capture progression patterns.

The output vector from the LSTM is then passed through fully connected hidden layers with dropout applied to mitigate potential issues such as gradient vanishing and overfitting. Finally, since the goal of the model is to differentiate AD from other causes of dementia, the output layer is a 2-dimensional vector representing the predictive logits for the True/False labels in this binary classification task. The overall model structure is shown in Figure 2.

We implemented a single layer LSTM (input dim=22, hidden dim=32) followed by a fully connected layer (dim=32) with ReLU activation and dropout (p=0.3). The model was trained using Adam optimizer (lr=0.001)

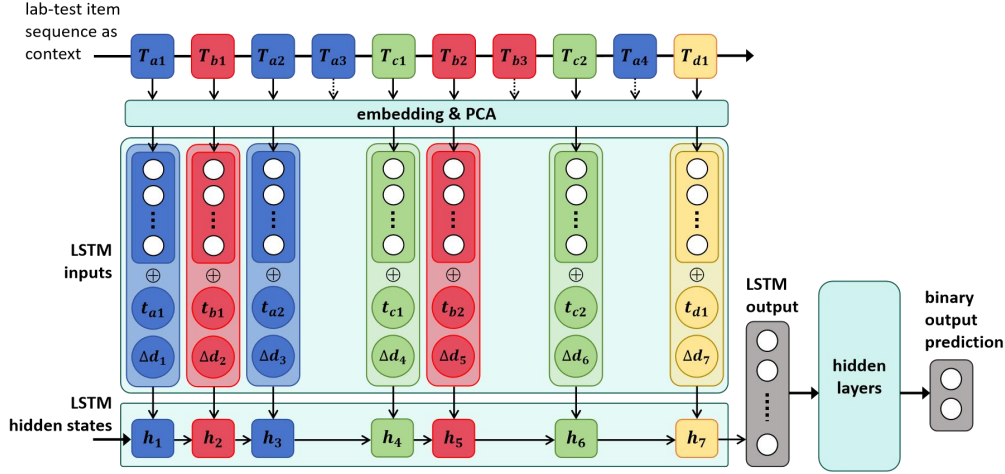

Figure 2: **Overall structure of the model.** $T$ denotes the lab-test items, $a, b, c, d$ and the corresponding colors represent different lab-test categories, namely the itemids, with the numbers following them indicating the orders within each category. $t_{kj}$ represents the test value, and $\Delta d_i$ denotes the scaled time gap. The symbol $\oplus$ represents the concatenation of vectors.

with cross-entropy loss.

## 2.6. Evaluation

The performance of different models and configurations will be evaluated by accuracy, precision, recall, F1-score, ROC curve, and AUC value[23]. The training set contains 3580 sequences from MIMIC-IV labeled as False (no AD), and 2510 sequences from both MIMIC-III and IV labeled as True (AD) with length $\geq 4$ constructed the training set, while 10% from each class constituted the test sets. Data samples from different time periods are further filtered when testing model performances with different time length before the diagnosis. The model performances are calculated by average values of metrics after running with 10 different random seeds.

## 3. Results

### 3.1. Embedding results

Given that the vectors are embedded from a comprehensive context from approximately 4000 patients, we intuitively believe that they have captured some contextual information already. To better illustrate this idea, the PC-processed vectors are projected into a 3-dimensional space by t-SNE algorithm for visualization. Though the space dimension was decreased by 17, the 3D vectors still show clear clusters for semantically similar lab-tests.

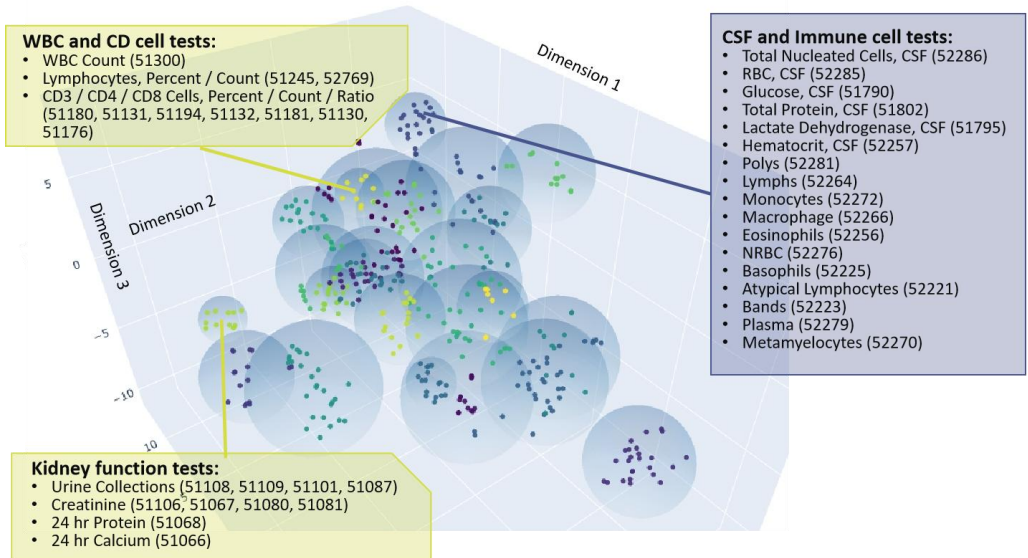

Figure 3: **Visualization of the lab-test items' with a 3-dimensional mapping.** The shaded spheres are generated using k-means clustering to highlight groups of tests that frequently co-occur within the same contexts.

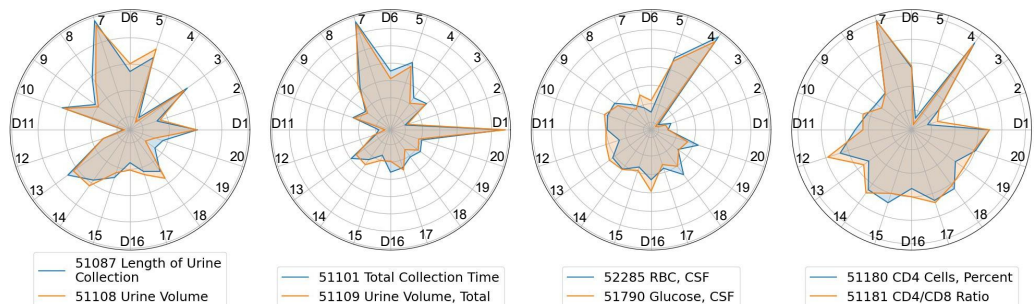

Figure 4: **Visualization of 20-dimensional vectors of** itemid**s in radar charts.** Lab-test items that usually appear together in patients' lab-test histories have very similar radar shapes, i.e., represented vectors.

As illustrated in Figure 3, a simple k-means clustering[24] with $k = 24$ was implemented to generate the shaded spheres for better visualization of the clusters. By analyzing the grouped items within each sphere, we found that those in the same sphere tend to co-occur more frequently within the same context compared to items from other groups, reflecting stronger contextual associations. This co-occurrence may stem from two primary factors: similarity and contextual relevance. Similarity refers to inherent characteristics shared by the items, such as their clinical functions, diagnostic purposes, or underlying biological processes. Contextual relevance, on the other hand, reflects how these items are commonly observed together within specific patient histories or medical scenarios, indicating their practical or situational associations in real-world applications. Three examples of clusters are given in the graph, showing the lab-test items' itemids and labels in the three chosen clusters. For instance, the Cerebrospinal fluid (CSF) tests which are used for diagnosing AD are gathered together. Vectors of four pairs of close items are further visualized in Figure 4, demonstrating that the embedding vectors effectively capture co-occurrence information from the historical sequences.

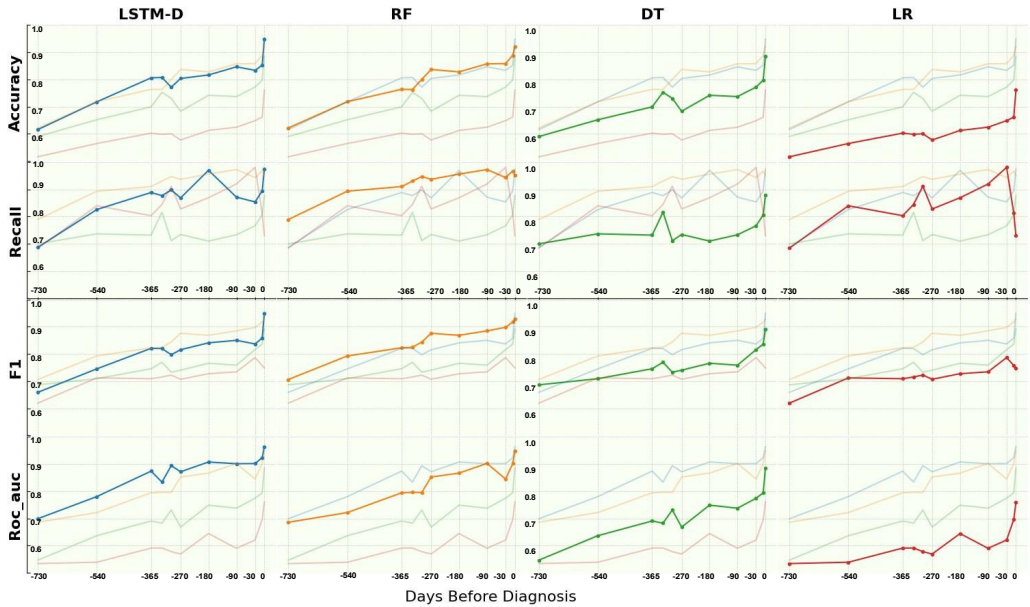

Figure 5: **Performances of different models.** In this 4 × 4 graph, each row denotes one metric, and each column represents one model. The x-axis denotes the number of days before diagnosis, covering a range of up to 2 years (730 days). 0 represents the diagnosis date, while negative values indicate days earlier than that. The y-axis denote the scores of different models measured in different metrics.

Table 3: Comparison of models' performances in diagnosis and early detection tasks. The diagnosis task was implemented with all the lab-test histories recorded in the EHR, while the early-detection task considered only the general tests (blood tests specifically) and the results recorded earlier than one year from the diagnosis admission.

|  | Early Detection ($S_E \otimes T_G$) - 1 Year | | | | | Diagnosis ($S_C \otimes (T_G + T_S)$) | | | | |
|---|---|---|---|---|---|---|---|---|---|---|
| Method | Acc | Prec | Rec | F1 | AUC | Acc | Prec | Rec | F1 | AUC |
| LR-Lasso | 0.6022 | 0.6338 | 0.8036 | 0.7087 | 0.5907 | 0.7607 | 0.7682 | 0.7296 | 0.7484 | 0.7600 |
| DT-Lasso | 0.6989 | <u>0.7593</u> | 0.7321 | 0.7455 | 0.6904 | 0.8849 | 0.9065 | 0.8780 | 0.8920 | 0.8855 |
| RF-Lasso | <u>0.7634</u> | 0.7500 | **0.9107** | **0.8226** | <u>0.7941</u> | 0.9208 | 0.9070 | 0.9512 | 0.9286 | 0.9475 |
| LSTM-D | **0.8056** | **0.7619** | <u>0.8889</u> | <u>0.8205</u> | **0.8738** | **0.9480** | **0.9252** | **0.9749** | **0.9494** | **0.9637** |

*Acc: Accuracy, Prec: Precision, Rec: Recall, F1: F1-score, AUC: Area Under the ROC Curve.*

## 3.2. Classification results

As described in the Methods section, we define the stages as the early stage $S_E$ and the complete stage $S_C$, as well as separating the tests into general tests $T_G$ and special tests $T_S$. We tested the models' performances with different time gaps, measured in days, before diagnosis for the early-detection tasks. The scores of various models including our model (LSTM-D), Random Forest (RF), Decision Tree (DT), and Logistic Regression (LR), are visualized in Figure 5. In this early disease detection task, minimizing false negatives is particularly crucial, as missing a diagnosis can have significant consequences for timely patient intervention and care. Consequently, recall, which measures the model's sensitivity in identifying the positive cases, is emphasized over precision. Some detailed values of the results are shown in Table 3, comparing the performance of the models in the diagnosis and early detection tasks. Here, we present the output from 1 year (365 days) before the diagnosis as a representative example of the early-detections.

In the early detection task, our model demonstrates high accuracy during the period of more than one year before diagnosis, matching the RF performance. Within the year before diagnosis, RF slightly outperforms LSTM-D in accuracy. Furthermore, RF shows high and stable recall and F1 scores, while our model, although performing better than DT and LR in these metrics, still has slightly lower performance than RF. Our model also maintains a high and stable AUC, consistently outperforming RF in early detection up to six months (180 days) before diagnosis.

Experiments with kernel-PCA were conducted as part of the dimensionality reduction strategy. For instance, kernel PCA and PCA combined to 20 dimensions yielded embeddings that, when trained with LSTM-D, achieved an accuracy of 0.8241 for one-year data and 0.9247 for all-time span data, with F1 scores reaching 0.8119 and 0.9281, respectively. A higher-dimensional configuration 40 dimensions further improved F1 scores to 0.8403 (one-year data) and 0.9299 (all-time span data). However, the results were not consistent, potentially due to the sensitivity of kernel PCA to hyperparameters and data variation, which led to its exclusion from the final approach.

In the diagnostic task (on the date of diagnosis), both LSTM-D and RF perform similarly well, achieving approximately 0.95 accuracy and AUC, along with recall scores greater than 0.95.

## 4. Discussion and Future Work

### 4.1. Our Model

#### 4.1.1. Pros and Cons

Our approach focuses on overcoming data heterogeneity, enabling it to incorporate almost all features and patients into consideration and capture hidden temporal information. This can help find potentially useful information for diagnosis that may not yet have been identified, demonstrating the value of temporal information for disease prevention and diagnosis. In contrast, traditional feature-based models need to rely on feature selection to perform well, and a large number of missing values and complex nonlinear relationships between features can lead to poor performance in linear models (e.g., DT, LR). However, although our model has shown competitive performance with RF, RF has a lower training cost and better interpretability. RF can capture non-linear information and thus handle data heterogeneity to some extent. In addition, RF can produce good performance with significantly shorter training times than deep networks. Therefore, despite our model's strong performance in metrics, RF is still preferred in practical scenarios.

#### 4.1.2. LSTM as Representative

One key aspect we aim to emphasize in this study is the proposition and validation of an often-overlooked piece of information: that the historical sequences of lab tests are valuable for the early detection and diagnosis of Alzheimer's Disease (AD). The focus of this research is not on designing a perfect time-series model. Therefore, we chose to use the relatively traditional yet structurally simple and interpretable LSTM model for time series analysis, rather than conducting exhaustive comparisons across various time-series models for minor performance improvements.

*4.1.3. Performances*

There is a significant accuracy and AUC decrease of LSTM-D at 300 days before diagnosis in Figure 5. The baseline models, including RF, also show different degrees of fluctuations at this time point, leading us to believe that this issue is not primarily due to our model design or implementation. This may be caused by the limitations of the data set, or it could be that AD progression at this stage exhibits features that are very similar to other types of dementia. This hypothesis has potential research value and could be tested by obtaining additional datasets to validate these assumptions.

*4.1.4. Hyperparameters*

Though we have tried several groups of hyperparameters, the model could be further improved by testing other combinations of hyperparameters more comprehensively. Some vibrations shown in Figure 5 might be caused by imperfect hyperparameters.

*4.2. Dataset*

MIMIC database only contains data from ICU patients, which might not be perfect for our study. In addition, this data set is very limited in size, especially for early detection tasks, which require a long period before diagnosis. A statistics of available AD patients' data is shown in Figure 6, illustrating that the available dataset for stages earlier than 1 month before diagnosis may be insufficient for a reliable model output. To address this challenge, we need a larger dataset that either for general hospitalization or longitudinal study of AD patients.

*4.3. Vector Representation*

*4.3.1. Embedding Approach*

We applied both PCA and kernel PCA for dimensionality reduction, with 20-dimensional embeddings for lab-test items, which has been proved useful for the model. We have tested some other numbers (e.g. 19) which did not show significant improvement. However, kernel PCA combined with PCA provided some improvements, particularly when increasing the dimensionality to 40 dimensions, as seen in the F1 scores for both one-year and all-time span data. However, kernel PCA showed inconsistency across different runs, likely due to its sensitivity to hyperparameters and data variation, which impacted the results' stability. Further experimentation with different kernel functions or the optimum number as well as adding additional features such as prescription patterns could be further explored.

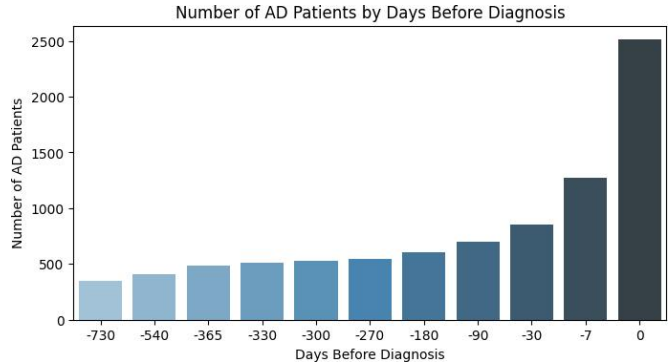

Figure 6: **Number of available patient data with respect to days before diagnosis.** The amount of available data on Alzheimer's disease (AD) patients decreases rapidly to less than 1000 within 1 month before diagnosis.

*4.3.2. Delta-time Representation*

The current temporal feature $\Delta d = \log(\text{time}(\text{diag}) - \text{time}(T))$ captures limited clinical semantic differences. Model performance could be improved through enhanced temporal representations or attention-based approaches with task-specific parameters and prior knowledge integration.

*4.3.3. Lab-test Value Scaling*

Since there were many different lab-test items with different ranges, it was hard to apply scalers for them respectively. Though it requires a lot of work, scaling the values properly may potentially enhance the performance.

## 5. Conclusion

In this study, we clarified definitions of diagnosis vs. early detection and general vs. special tests. Then, we proposed novel approaches for embedding lab-test data, and for early detection and differential diagnosis of dementia using sequential information from general lab tests. By drawing an analogy between medical test sequences and natural language, we applied word2vec embedding to capture the intrinsic semantic relationships between lab tests, resulting in meaningful clusters of related tests in the embedding space. Our LSTM model not only effectively processes the temporal features of test sequences but also achieves promising performance in both early detection and diagnostic tasks using only general lab data, demonstrating the potential of identifying early-stage dementia through daily test data.

However, limitations remain, including the use of linear dimensionality reduction, temporal feature design expressiveness, and potential dataset selection bias in the current dataset. Future research directions worth exploring include investigating nonlinear dimensionality reduction techniques, more sophisticated temporal modeling methods, and validation on more diverse non-ICU patient populations.

## 6. Acknowledgment

*6.1. Author Contribution Statement*

Yizong Xing: Conceptualization, Methodology, Software, Validation, Formal Analysis, Data Curation, Visualization, Writing - Original Draft, Writing - Review & Editing. Dhita Putri Pratama: Methodology, Investigation, Validation, Writing - Review & Editing. Yufan Zhang: Investigation, Methodology, Validation, Writing - Review & Editing. Yuke Wang: Investigation, Methodology, Validation, Writing - Review & Editing. Brian E. Chapman: Supervision, Resources, Writing - Review & Editing.

**Declaration of Generative AI and AI-Assisted Technologies**

During the preparation of this work, the authors used Writefull in order to improve language. After using this tool, the authors reviewed and edited the content as needed and take full responsibility for the content of the publication.